# Exploring the Emerging Technologies within the Blockchain Landscape

Mohammad Ali Tareq [1[0000-0001-5521-8814]], Piyush Tripathi [2[0009-0007-1798-5941]], Nurhayati Md Issa [3], Mahdi H. Miraz [4,5,6,*[ 0000-0002-6795-7048]]

[1] Universiti Malaya (UM), Malaysia
[2] Texas A&M University, College Station, USA
[3] Yamaguchi University, Japan
[4] Xiamen University Malaysia, Malaysia
[5] Wrexham University, UK
[6] University of South Wales, UK
m.miraz@ieee.org

**Abstract.** Although blockchain technology was first introduced in 2008 and materialised in 2009, the early usage of blockchain were mainly limited to financial technologies, particularly cryptocurrencies. Later, blockchain became a widespread emerging technology, utilised in multifaceted sectors and applications. In fact, various new and innovative application of blockchain and distributed ledger technologies are still continuously being researched and explored. On the other hand, smart-contracts were first introduced in 1990s, however, it did not gain enough popularity until being integrated with blockchain technologies lately. The duo lately been seen as the key to many innovations in various industries and sectors. So, we took data from 1445 blockchain-related patent documents and tried to map out the historical and current trends in patenting activities in the blockchain field. This helps us get a better grasp of how blockchain technologies are evolving and being tracked. In addition to serving as an indicator of science and technology growth, patents are also used to judge the research potential and development of a particular technology.

## 1. Introduction

Szabo, an American lawyer as well as computer scientist, was the first to conceptualise smart-contracts as far back as the early 1990s [1]. Obviously, his version of smart-contracts was in the absence of today's blockchain or distributed ledger technologies. As a matter of fact, the idea of blockchain was first put forward by [2] in 2008 as an incidental result of the Bitcoin cryptocurrency, materialised in 2009. The earliest version of the blockchain, i.e. Blockchain 1.0 [3], was in fact orchestrated without the compatibility of smart contract. The fusion of both the blockchain and the smart-contracts was first transpired in Blockchain 2.0 [3]. Since then, the application of this duo has been exponentially increasing in various technological innovations beyond cryptocurrencies [4], as evident in various research publications and patent applications. Such implementation of smart contracts on blockchain' (distributed) ledgers not only assumes it immutability, transparency and extra layer of security, but also codify and automate the 'delivery' aspect of a contact, enabling secure and automated execution of any part of a project upon satisfaction of the predefined conditions, being verified by the peers of the distributed blockchain network utilising its consensus model. That being the case, such fusion of the blockchain and smart-contract technologies has been acting as a spring-board for multitude of innovations in multifaceted disciplines. Therefore, to study the current trends and predict the future directions of technological innovations applying blockchain and smart-contracts, we have used the IPC Codes in this study to analyse subject trends of the blockchain related patents. This study tries to identify the technologies pertaining to the blockchain technology and further exploring the relationship among the technologies used in the development of blockchain. Our investigation utilizes this information to tackle our study query: which fields of knowledge play a role? The study is the initial step to map the overall technology space within the blockchain. This study will lay the groundwork by addressing the core technologies as well as the interrelationship among them, and also the assignees using these technologies.

The rest of this paper is as follows: literatures are discussed in Section 2. Section 3 explained the methodology of the paper. Findings are presented in Section 4 and conclusion follows in Section 5.

## 2.   Literature Review

### 2.1.   Blockchain Technologies

A distributed ledger of transactions, or blockchain, is an developing technology that can be traced back to 2008 when [2] introduced bitcoin [5-9]. Unlike traditional databases, blockchains are made up of blocks containing transactions validated by encryption mechanisms, and their information cannot be removed or altered by anyone. These blocks are consistently linked, resulting in an ever-expanding chain of blocks [10]. A blockchain technology is generally viewed as a cross between software engineering, cryptography, economic game theory, and decentralized computing.

Attention to cryptocurrencies has increased 2014 onward, and approximately 9,984 cryptocurrencies are listed in CoinMarketCap upto July 2023. A blockchain can be used for more than just cryptocurrencies. In today's world, blockchain technology is being used in a wide range of domains, including financial services [11-13], smart contracts [14], Internet of Things (IoT) [15, 16], identification services [9], security services [17], healthcare services [18-20], supply chain [21-25], etc. Diverse focus of blockchain technology warrants a proper technology mapping for a better understanding of the evolution in this field. Fig. 1 shows a variety of blockchain applications in both financial and non-financial domains, as adopted from Casino, Dasaklis [26].

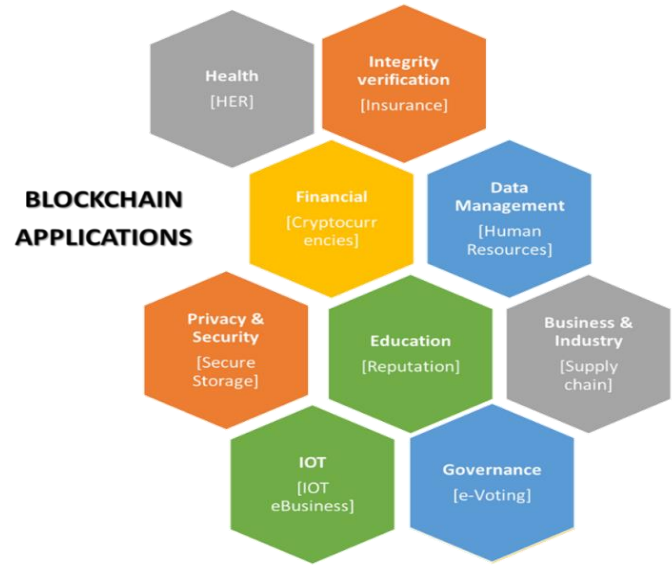

**Fig. 1.** Classification of blockchain applications by Casino [26]

### 2.2. Patent Analysis

The World Intellectual Property Organization (WIPO) defines a patent as a documentation showcasing a unique product, a pioneering procedure for product creation, or an inventive answer to a specific issue. This document grants inventors the exclusive rights and legal protection for a specified period, playing a crucial role in promoting equitable technology development and distribution [27, 28]. A patent document typically includes essential technological details such as the patent number, issuance date, inventor's name, assignee (the entity granted the patent rights), abstract, title, and an International Patent Classification (IPC) code that categorizes the technology's subject matter.

Patents serve as a crucial gauge of innovation, assessing the advancement of technology, inventive endeavours, and creative accomplishments related to novel technologies, products, and processes. Utilising patent intelligence offers a valuable means of comprehending a nation's knowledge-driven economy. Despite their limitations, patent indicators remain an unparalleled source for analysing the strides made in technology innovation.

By providing detailed technical information, patents offer new knowledge and insights. For nearly two decades, experts and researchers in technology management and technology forecasting have employed the analysis of patent documents as a method to identify and understand the technologies associated with various industries [28-33]. Analysing patents helps prevent researchers from duplicating efforts in specialized fields and empowers them to create novel technological advancements based on existing information. This approach has proven valuable in gaining insights into technological trends and advancements within specific sectors. In essence, a patent document serves as an extensive repository of technological information that can be valuable for research and innovation purposes.

### 2.3. Patent management

Traditional patent management encompasses various activities such as applying for patents, securing their issuance, obtaining them, ensuring their maintenance, and safeguarding them against infringement. This process necessitates a combination of robust technical and legal expertise. Effective patent management enables researchers and firms to gauge the efficacy of their R&D activities, aiding in strategic planning. It also serves as a means to gain a competitive advantage concerning intellectual property rights within the industry, setting them apart from their rivals.

To devise a successful R&D strategy, careful consideration of future technology trends is essential. Analysing patents can serve as a valuable source of technology forecasting, providing insights into emerging technologies and potential areas for investment and development. For any firm, introducing a novel technology is of paramount importance, as it can significantly enhance competitiveness in the market. By continuously innovating and leveraging unique technologies, companies can position themselves as industry leaders and gain a distinctive edge over competitors.

Compared to scholarly articles, patents offer more comprehensive details on various technologies. Examining the data gleaned from patent documents has given rise to strategic organization, handling of technology, analysis of business rivals, and managing research and development divisions [28, 29, 34]. Analysing patent data has been instrumental in strategic planning, as it allows organizations to identify technological opportunities, potential areas of growth, and emerging trends. It aids in effective technology management, helping companies align their R&D efforts with market demands and innovation goals. Besides, through competitor analysis, organizations can better understand the technological landscape and the strengths and weaknesses of their rivals, enabling them to make informed decisions in the market [29]. An analysis of competitors' patents and a new exploration of technology opportunities can be achieved from patents' bibliographic information such as inventors, publication dates, applicants, and other specifications [29, 32, 35, 36]. Analyzing patents can be advantageous in predicting technology trends and forecasting, crafting technology roadmaps, and identifying countries that are pioneers in innovative technology [29].

### 2.4. R&D, Technology and IPC

Researchers, investors, and public agencies are taking a close look at blockchain technology, which has a variety of impressive applications. A good way to make an informed technology investment and selection decision is to recognize the future trend of technologies. It is the potential for technological advancement in general or within a particular field that affects the overall industry and individual companies. Different industries' R&D intensity is affected by different technological opportunities, which leads to heterogeneous R&D productivity. Different enterprises have different R&D productivity and operating results in response to technological opportunities at the firm level.

The International Patent Classification (IPC) is the world's main system for categorizing patents, as recommended by the World Intellectual Property Organization (WIPO). Each IPC comes with a matching technology category. By evaluating IPC classification distributions, it is possible to assess the overall technology trend by analysing the technology classification distributions of patents.

IPC codes can be used to pre-defined technologies and can be used to investigate their trends, i.e.; the sequence of technology definition. In investigating IT and BT areas, [37] used IPC codes to assign relevant patents to information technology (IT) and biotechnology (BT), and analysed the characteristics of patents in each of the IPC codes to identify emerging technologies by defining one IPC code as one technology.

The Strasbourg Agreement of 1971 initiated the establishment of an ordered patent classification known as the IPC [23]. The International Patent Conversation (IPC) codes have been overseen by the World Intellectual Property Organization (WIPO). These codes function as a technological categorization term that gives testament to the subject of the invention. In this paper, we use the 4th level of IPC code such as G10D 1/02 as in Table 1:

**Table 1.** Hierarchical structure of IPC codes

| IPC Hierarchy | | Description |
|---|---|---|
| **G** | : | Section G - Physics |
| **G10** | : | Musical Instruments; Acoustics |
| **G10D** | : | Stringed musical instruments; Wind-Actuated musical instruments; Accordions or concertinas; Musical Instruments not otherwise provided for; |
| **G10D 1/02** | : | of violins, violas, violoncellos, basses |

**2.4.1 Necessity of Technology Analysis**

The trends in research and development (R&D) paradigms highlight the importance of dominant designs for shaping the development of next-generation products and the significance of discontinuous innovation for creating breakthrough products [38]. Innovative ideas can originate from various sources, and while internal sources have traditionally been considered the primary suppliers of new ideas, recent studies suggest that customers and users play a significant role in generating a majority of ideas in many industries. Moreover, suppliers and even competitors can also offer valuable insights that contribute to the innovation process. Discontinuous innovation involves radical shifts and ground-breaking developments that result in entirely new products or solutions. These innovations disrupt the existing market and can revolutionize entire industries. The Blockchain is regarded as one of the emerging technologies that has the potential to disrupt the entire ecosystems in different areas where blockchain is involved in as shown by [26]. This has opened up the necessity to explore the technology development within the blockchain ecosystems. The escalating significance of blockchain technology and its promising future has prompted key research to examine patents within this sector [39].

What then becomes crucial is how to identify these newer innovations that transform raw data into information that can be used for innovation, regardless of source. From macrolevel analysis of strategy to microlevel modelling of specific emerging technologies, patents are useful sources of information about technical progress and innovative activity [30, 33, 40]. Analysing patents can reveal technological details, reveal business trends, inspire new industrial solutions, and guide investment decisions [27, 31, 34, 41].

## 3. Methodology

The prime objective of this paper is to explore the technology landscape of patents related to blockchain. The sample of this paper consists of 1,445 patents from 2020 and 2021. This sample will allow is to have an idea on the current development as well as the impact of these technologies. The patents were handpicked after careful review of the abstracts and the claims in order to avoid auto-generated samples containing near keywords 'chain-saw', 'building block' and alike. After curation and sorting of the sample, necessary information (Publication number, Dates, Assignee, Country, Abstracts and IPC codes) from 1,426 patents are collected for further analysis. Excel, Python and Gephi are used to analyse the curated and sorted data and to create graphical presentations. Figure 2 presents the flow of methodology for the paper.

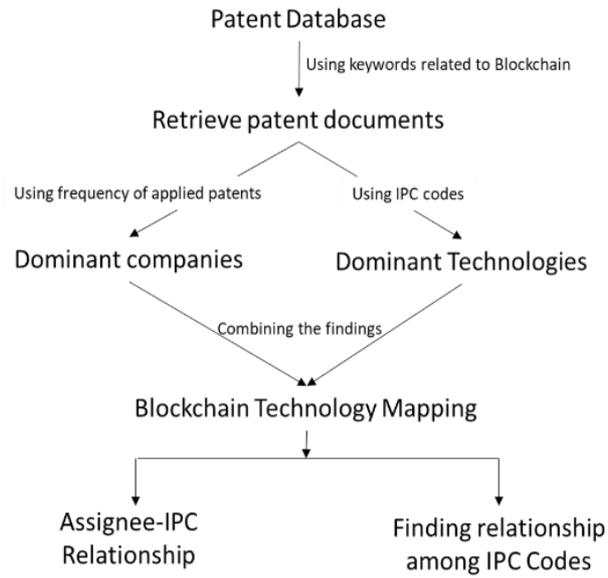

**Fig. 2.** The research flow

## 4. Findings

Figure 3 shows the trends in patenting activities for blockchain related patents documents over the years. The spike in the blockchain patents started from around 2016 with an increasing trend. Interesting to see the decline in the number of application post-2021.

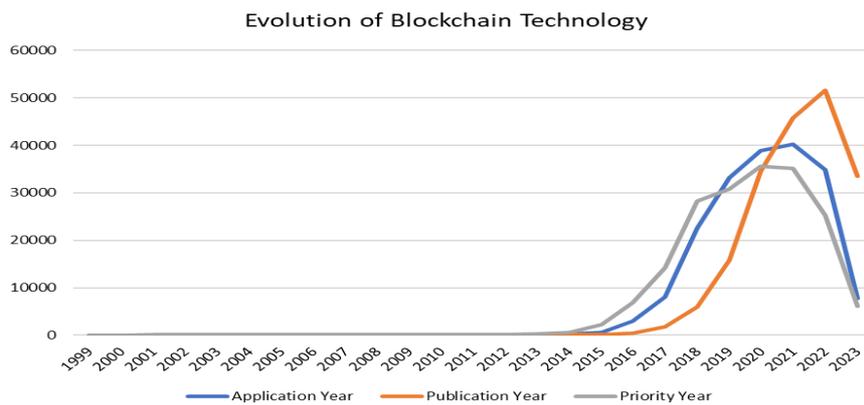

**Fig. 3.** Trend in patenting activities

The top owners of patents in the blockchain field are the E-commerce giant ALIBABA, Chinese Technology Company Tencent, IBM technology firm, Ping An tech, Nchain Holding, Advanced New Technologies, and Alipay finance services. Surprisingly, majority of the assignees of the patents are from China. The Non-Chinese firms are IBM, Siemens and Microsoft among the Top 20 as shown in figure 4.

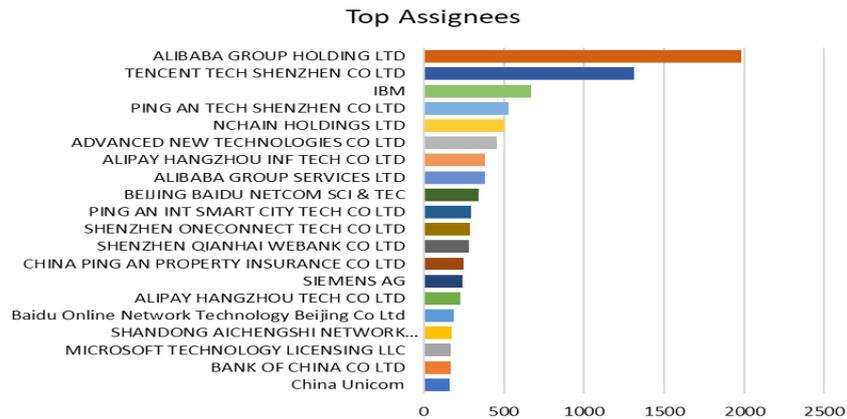

**Fig. 4.** Top assigners

Patents are organized based on the International Patent Classification (IPC), a universal and standardized system provided by WIPO, which enables fast assessment and search. For the sample of 1,426, total of 61 major IPC classifications are covered by the blockchain related patents. As perceived, the technology related to the computing and electric communication technique are the most focused ones. Interesting to see Health Informatics related technologies (G16H) has started to be focused more recently. The list of Top 10 IPC Subclasses has been represented in Table 2.

**Table 2.** Top 10 IPC subclasses.

| IPC Subclass | Descriptions |
|---|---|
| G06Q | Data Processing Systems or Method's |
| H04L | Transmission of Digital Information, e.g. Telegraphic Communication |
| G06F | Electric digital data processing |
| G06K | Recognition of Data; Presentation of Data, Record Carries, Handling Records Carriers |
| H04W | Wireless communication networks |
| G06N | Computer systems based on specific computational models |

| | | |
|---|---|---|
| G16H | Healthcare Informatics, i.e. Information and Communication Technology [ICT] Specially Adapted for the Handing or Processing of Medical or Healthcare Data | |
| G07C | The time or Attendance Registers; Registering or Indicating the Working of Machines; Generating Random Numbers; Voting or Lottery Apparatus | |
| G07F | Coin-Freed or Like Apparatus | |
| H04N | Pictorial Communication, e.g. Television | |

We further investigated whether the top owners are having diverged technology focus or not. The highest IPC subclass used was by Koninklijke Philips N.V. (Patent No. US10811771B1) with 9 different occurrences; followed by Strong Force Tx Portfolio 2018 LLC (US20200272469A1) and 2 more patents from Koninklijke Philips N.V. Lbxc Co Ltd as well as Rolls Royce Holdings Plc follow thereafter. Surprisingly, none of the companies on the list in the Table 3 are among the Top assignees of the blockchain related patents. Detailed analysis in the future can give a clearer picture of the technology adoption by the companies.

**Table 3.** Companies with diverged technology focus

| No | Publication Number | Assignees | G06Q | H04L | G06F | G06K | H04W | G06N | G16H | G07C | G07F | H04N | Total IPC Subclasses |
|---|---|---|---|---|---|---|---|---|---|---|---|---|---|
| 1 | US10811771B1 | Koninklijke Philips N.V. | | 1 | | 1 | 1 | 1 | | | | | 9 |
| 2 | US20200272469A1 | Strong Force Tx Portfolio 2018 Llc | 1 | 1 | 1 | 1 | | 1 | | | | | 7 |
| 3 | US20200364187A1 | Koninklijke Philips N.V. | 1 | | 1 | | 1 | 1 | | | | | 6 |
| 4 | US20200358183A1 | Koninklijke Philips N.V. | | | | | 1 | 1 | | | | | 5 |
| 5 | WO2020209413A1 | Lbxc Co Ltd | 1 | 1 | | | 1 | 1 | | | | | 5 |
| 6 | US20200193464A1 | Rolls Royce Holdings Plc | 1 | 1 | | 1 | 1 | | | | | | 5 |
| 7 | US10733160B1 | State Farm Mutual Automobile Insurance Co. | 1 | 1 | 1 | | | | | 1 | | | 5 |
| 8 | US20200233398A1 | General Electric Company | | 1 | 1 | | | | | | | | 5 |
| 9 | KR2141219B1 | Suh Dong Geun & Seo Jeong-Seong | 1 | | 1 | | | 1 | | | | | 5 |
| 10 | KR2149245B1 | Innodigital Co Ltd & Hanshin University | 1 | 1 | 1 | | 1 | | | | | | 5 |
| 11 | EP3761255A1 | Abb Asea Brown Boveri Ltd | 1 | 1 | | | | | | | 1 | | 5 |
| 12 | CN111667318A | Guangzhou 9skychina Information Technolo | 1 | 1 | 1 | 1 | | 1 | | | | | 5 |
| 13 | CN111476656A | Shenzhen Zhaji Network Technology Co Ltd | 1 | 1 | 1 | | | | | | 1 | 1 | 5 |

Figure 5 presents the occurrences of the technologies within the sample for this study. These IPCs are the 4th level of the IPC categories which can give a detailed description of the major technologies needed for the efficient and effective operation of blockchain.

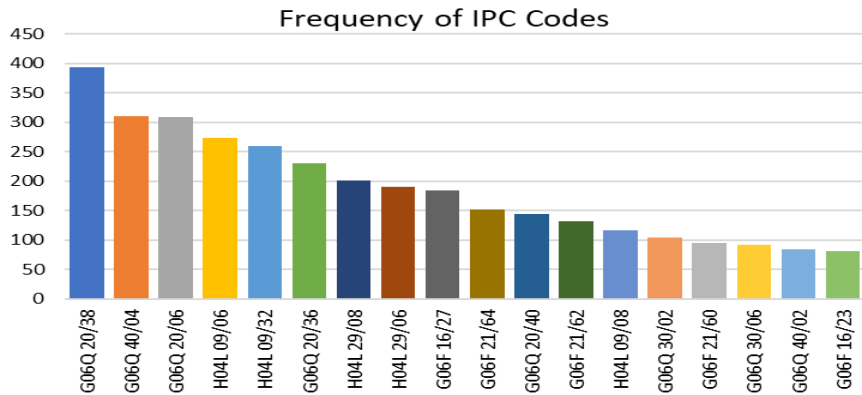

**Fig. 5.** displays the contributions made by each category in the IPC classification within the collected patents.

The network diagram in figure 6 is the most important presentation in the sense that it shows the interlinkages among the IPC Codes within the blockchain domain. Complementary technologies can be identified using this network. As perceived, the top technologies are found to be interlinked with each other.

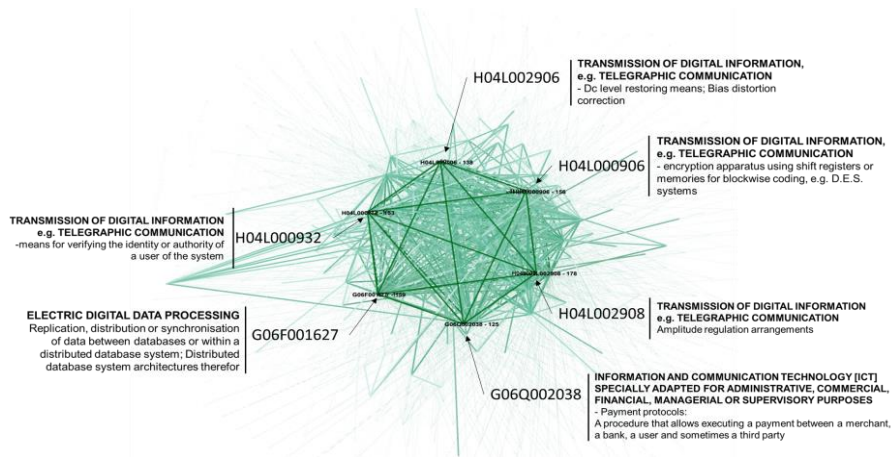

**Fig. 6.** IPC Codes within the blockchain domain.

Figure 7 presents the distribution of the 4th level IPC Codes within the respective top assignees. This figure can help the analyst to identify the competitors and collaborator within the same technology fields. Further analysis of this details can provide an idea on the possible M&A activities among the companies for gaining competitive advantages.

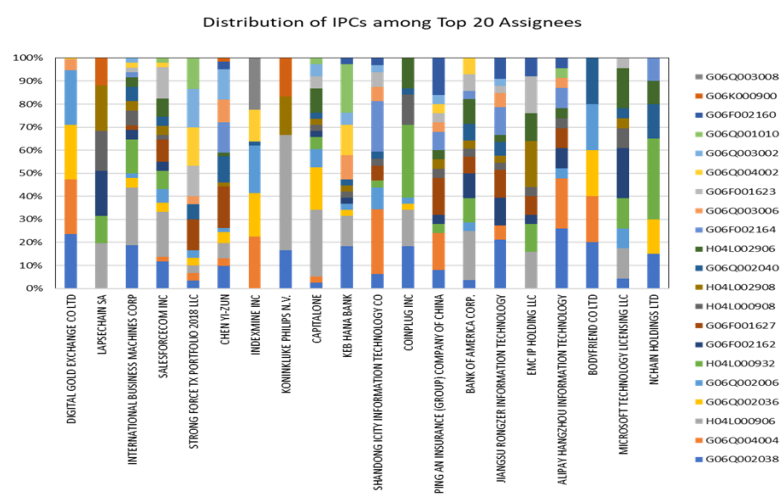

**Fig. 7.** The distribution of the 4$^{th}$ level IPC.

## 5. Conclusion

This paper analysed 1,426 patents related to blockchain to explore the technology focus within the blockchain landscape. One of the interesting findings, among others, is that the diverged technology focused companies are not among the top assignees, and findings of the paper can be helpful for potential M&A activities leading to competitive advantage. This paper is a pilot study for an extensive study and to develop technology map for the emerging area like blockchain.

## Acknowledgement

This work was funded by Xiamen University Malaysia Research Fund (XMUMRF) under Grant XMUMRF/2021-C8/IECE/0025 and XMUMRF/2022- C10/IECE/0043.